\documentclass[11pt, a4paper]{article}
\usepackage{moriond,epsfig}

\def\be{\begin{equation}}
\def\ee{\end{equation}}
\def\bea{\begin{eqnarray}}
\def\eea{\end{eqnarray}}

\begin{document}
\title{HIGHLIGHTS FROM THE NA60 EXPERIMENT}

\author{ A.~FERRETTI $^{(1)}$, R.~ARNALDI $^{(1)}$, R.~AVERBECK $^{(8)}$ , K.~BANICZ $^{(4)}$, J.~CASTOR $^{(3)}$, B.~CHAURAND $^{(6)}$, C.~CICAL\` O $^{(9)}$, A.~COLLA$^{(1)}$,$^{(2)}$, P.~CORTESE $^{(1)}$, S.~DAMJANOVI\v C $^{(4)}$,$^{(2)}$, A.~DAVID $^{(2)}$, A.~DE~FALCO $^{(9)}$, A.~DEVAUX $^{(3)}$ , A.~DREES $^{(8)}$, L.~DUCROUX $^{(10)}$, H.~EN'YO$^{(7)}$, M.~FLORIS$^{(9)}$, A.~F\" ORSTER$^{(2)}$, P.~FORCE$^{(3)}$, N.~GUETTET $^{(2)}$, A.~GUICHARD $^{(10)}$, H.~GULKANYAN $^{(11)}$, J.M.~HEUSER $^{(7)}$, M.~KEIL $^{(2)}$, L.~KLUBERG $^{(6)}$, C.~LOUREN\c CO $^{(2)}$, J.~LOZANO $^{(5)}$, F.~MANSO $^{(3)}$, P.~MARTINS $^{(5)}$, A.~MASONI $^{(9)}$, A.~NEVES $^{(5)}$, H.~OHNISHI $^{(7)}$, C.~OPPEDISANO $^{(1)}$, P.~PARRACHO $^{(5)}$, P.~PILLOT $^{(10)}$, G.~PUDDU $^{(9)}$, E.~RADERMACHER $^{(2)}$, P.~RAMALHETE $^{(5)}$, P.~ROSINSKY $^{(2)}$, E.~SCOMPARIN $^{(1)}$, J.~SEIXAS $^{(5)}$, S.~SERCI $^{(9)}$, R.~SHAHOYAN $^{(5)}$, P.~SONDEREGGER $^{(5)}$, H.J.~SPECHT $^{(4)}$, R.~TIEULENT $^{(10)}$,  G.~USAI $^{(9)}$, R.~VEENHOF $^{(2)}$ AND  H.~W\"OHRI $^{(9)}$}

\address{(1)Univ. di Torino and INFN, Torino, Italy. (2)CERN, Geneva, Switzerland. (3)LPC, Univ. Blaise Pascal and CNRS-IN2P3, Clermont-Ferrand, France. (4)Univ. Heidelberg, Heidelberg, Germany. (5)IST-CFTP, Lisbon, Portugal. (6)LLR, Ecole Polytechnique and CNRS-IN2P3, Palaiseau, France. (7)RIKEN, Wako, Saitama, Japan. (8)SUNY, Stony Brook, NY, USA. (9)Univ. di Cagliari and INFN, Cagliari, Italy. (10)IPN-Lyon, Univ. Claude Bernard Lyon-I and CNRS-IN2P3, Lyon, France. (11)YerPhI, Yerevan, Armenia.}

\maketitle\abstracts{The NA60 experiment is a fixed-target experiment at the CERN SPS. It has measured the dimuon yield in Indium--Indium collisions with an In beam of 158 AGeV/{\it c} and in p-A collisions with a proton beam of 400 and 158 AGeV/{\it c}. The results allow to address three important physics topics, namely the study of the $\rho$ spectral function in nuclear collisions, the clarification of the origin of the dimuon excess measured by NA50 in the intermediate mass range, and the J/$\psi$ suppression pattern in a collision system different from Pb-Pb. An overview of these results will be given in this paper.}

\noindent
{\it \bf The NA60 experiment.} The measurement of dimuon production is a key tool to gain insight into ultra-relativistic nuclear collisions. However, the importance of the dimuon data depends strongly on the resolution of the experimental apparatus. While the NA50 experiment measured the  J/$\psi$  suppression pattern as a function of centrality in Pb--Pb collisions, its mass and vertexing resolutions were not sufficient to address two other important physics topics, namely the shape of the in-medium spectral function of the $\rho$ meson and the origin of the dimuon excess observed by NA50 in the intermediate mass region (IMR, 1.2-2.7~GeV/$c^2$).

The NA60 experiment inherited from NA50 the muon spectrometer (MS) for muon triggering and tracking, and the Zero Degree Calorimeter (ZDC), which measures the energy carried forward (at $0^{\circ}$) by the spectator nucleons, to evaluate the centrality of A--A events. Two components were added: before the target, a beamscope (BS) made of two pairs of cryogenic silicon strip detectors determines the beam impact point on the target transverse plane with a 20~$\mu$m resolution; after the target, a vertex tracker (VT), made of 16 planes of radiation-hard silicon pixel detectors embedded in a 2.5~T dipolar magnetic field, measures tracks and their momenta before they suffer multiple scattering in the absorber (for details on the NA60 apparatus see~\cite{Usai:2005zh}).

Tracks in the VT are matched (both in momenta and in coordinate space) to tracks measured in the muon spectrometer. This procedure greatly improves the mass resolution, particularly in the low-mass range (20~MeV/$c^2$ at the $\omega$ mass, to be confronted with NA50 80~MeV/$c^2$). This allows to resolve and evaluate the different contributions to the dimuon spectrum below 1.2 GeV/$c$. Moreover, it is possible to determine the offset between muon tracks and the primary interaction vertex with a precision of $\sim 40~\mu m$  for 20~GeV/$c$ muons. This is sufficient to discriminate between muons coming directly from the fireball and muons originated by secondaries decays in the intermediate mass range. 

The results reported in this paper were obtained from the analysis of data taken in 2003 for In--In collisions at 158~AGeV/$c$, and in 2004 for p--A collisions at 158~GeV/$c$. The choice of a colliding system different from Pb--Pb makes it possible to search for the scaling variable which drives the onset of the J/$\psi$ anomalous suppression. Moreover, p--A data at 158~GeV allow to compare p--A to A--A data without systematic errors deriving from the energy rescaling.

\begin{figure}[!htb]
\resizebox{0.50\textwidth}{0.26\textheight}{%
\includegraphics*[bb = 0 0 560 688]{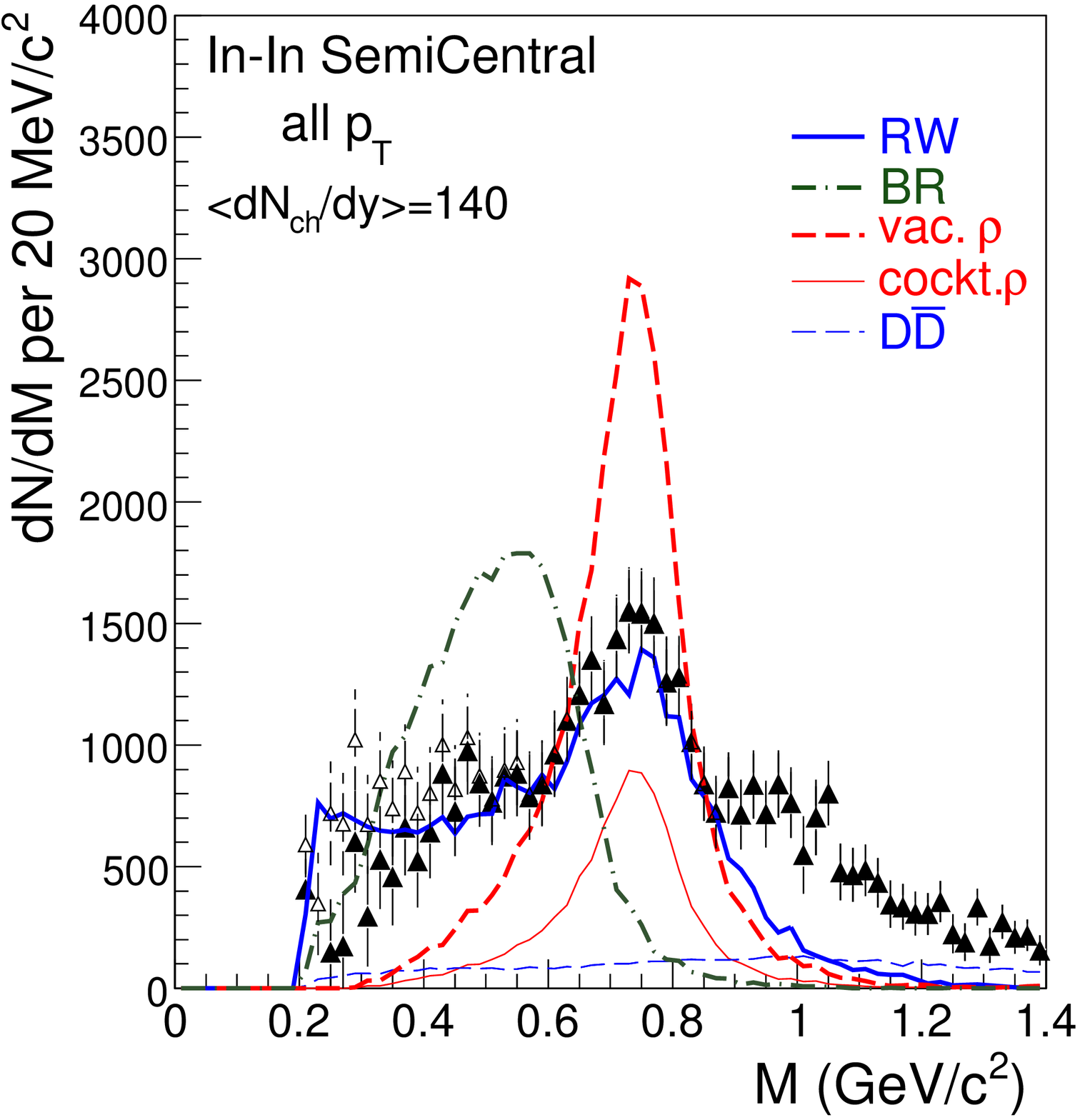}}
\resizebox{0.5\textwidth}{0.26\textheight}{%
\includegraphics*[bb = 0 0 560 620]{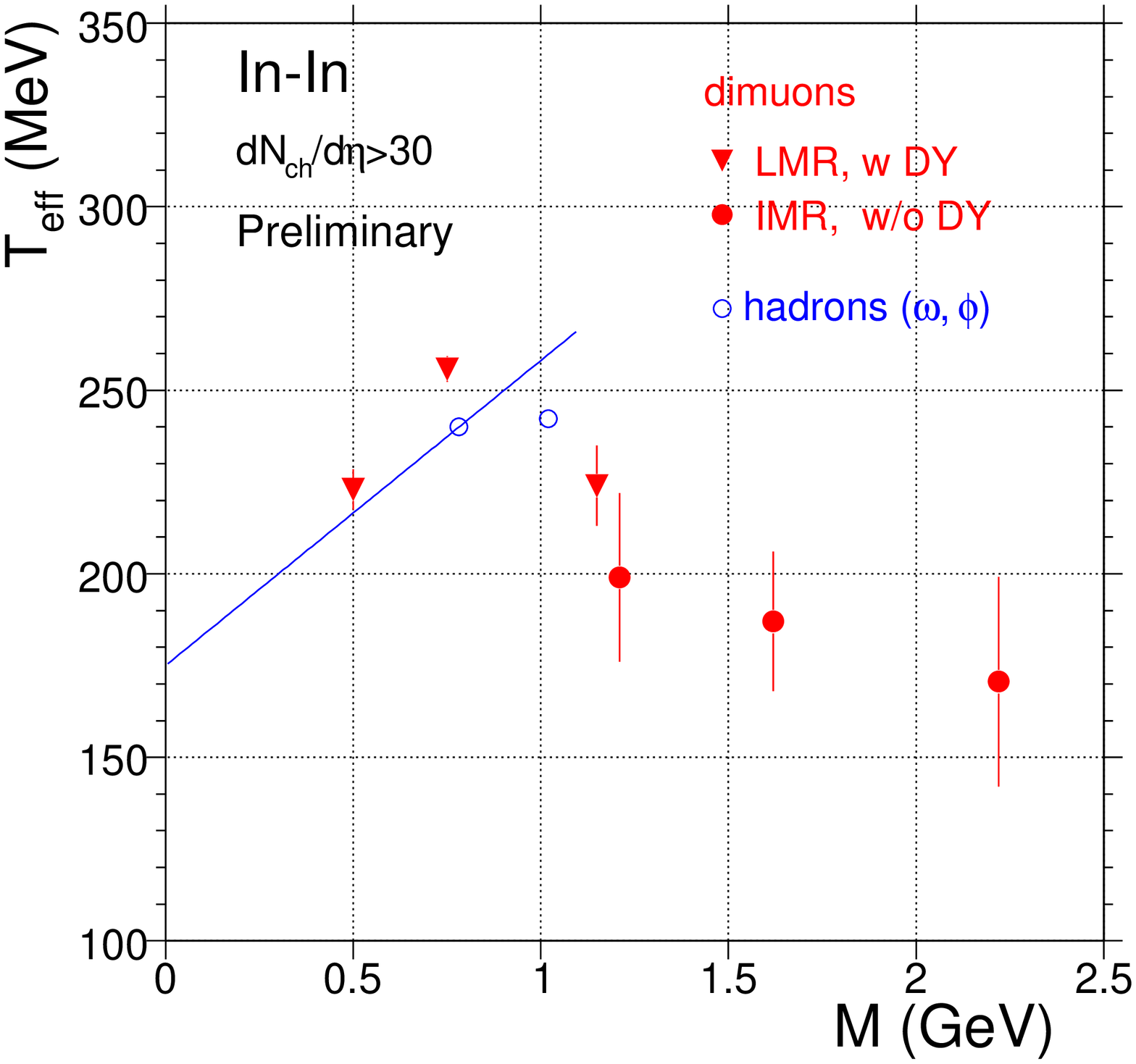}}
\vglue -4mm
\caption{Left: comparison of excess mass spectrum (black triangles) with no $p_T$ cut to model predictions made for In-In at $dN_{ch}/d\eta=140$ (semicentral bin). Cocktail $\rho$ (thin solid), unmodified $\rho$ (dashed), in-medium broadening $\rho$ \protect\cite{Rapp1}, in-medium moving $\rho$ related to \protect\cite{Brown} (dashed-dotted), uncorrelated charm (thin dashed). Errors are purely statistical. The open data points show the excess resulting from a decrease of the subtracted $\eta$ yield by $10\%$. Right: $T_{eff}$ from an exponential fit to the $p_T$ window $0.6-2$ GeV/$c$ vs. mass. The open circle points are the $\omega$ and $\phi$ temperatures. Full circle points are obtained from a different analysis of the IMR up to 2.5 GeV/$c^2$ (see fig.\protect\ref{fig2}). The line is the expected temperature on the basis of a linear dependence on meson mass as determined by the NA49 systematics \protect\cite{NA49systematics}.}
\label{fig1}
\end{figure}

{\it \bf The low mass region.} The net opposite-sign dimuon spectrum below 1.4 GeV/$c^2$ has been obtained from the raw mass spectrum by subtraction of combinatorial background and of signal fake matches between tracks in the MS and in the VT. Four centrality classes were defined via the charged particle multiplicity measured by the VT. Most peripheral data ($4< dN_{ch}/d \eta <30$) are well reproduced by the cocktail of expected electromagnetic decays of the neutral mesons, while for more central collisions a strong excess appears, whose shape is not known {\it a priori}. 
Thanks to the high data quality, it is possible to isolate this excess by subtracting from the data a hadron cocktail without the $\rho$ (see\cite{prl}), whose spectral shape is expected to be modified in the fireball. The resulting excess is shown in Fig.\ref{fig1} (left) for semicentral collisions ($110< dN_{ch}/d \eta <170$): it is characterized by a peaked structure centered on the nominal $\rho$ mass, which broadens and increases with centrality and resides on a wide continuum. However, a quantitative analysis of the excess (for details see~\cite{lowmassHP06}) demonstrates that the ratio between the continuum-subtracted peak and the $\rho$ obtained from the cocktail fit decreases by almost a factor 2 from most peripheral to most central bin. This means that the excess can not be simply interpreted as the cocktail $\rho$  on top of a continuum.



As shown in fig.\ref{fig1} (left), the moving mass model related to Brown/Rho (BR) scaling is ruled out. The qualitative features of the excess mass spectra are consistent with the interpretation as direct thermal radiation from the fireball, dominated by $\pi \pi$ annihilation. Models based on the in-medium $\rho$ broadening scenario and which take into account the role of baryons in the broadening \cite{Rapp1} are able to reproduce quantitatively the data below 0.9~GeV/{\it c}.
For the mass region above 0.9~GeV, data seem to be described equally well by introducing 4$\pi$ hadronic processes sensitive to vector-axialvector mixing (and therefore to chiral symmetry restoration) or in terms of partonic processes dominated by $q\overline q$ annihilation. This feature could be a manifestation of parton-hadron duality. 

The study of $p_T$ spectra presents further interesting features. The trend at small $p_T$ is opposite to the flattening expected from radial flow of hadrons produced at kinetic freeze-out, while it flattens as expected in the $\phi$ mass bin. Moreover, the $T_{eff}$, obtained from fits of the $m_T$ spectra (see fig.~\ref{fig1} right) and plotted as a function of mass, shows a maximum in the $\rho-$like region. This may be attributed to the $\rho$ produced at freeze-out, which experiences the largest blue-shift and thus the highest effective temperature. Besides that, it is worth noting that the continuum above 0.9 GeV/$c$ is cooler than the continuum below 0.6~GeV/$c$: this seems to indicate that the two regions are fed by qualitatively different sources. It is hoped that a finer theoretical understanding of $p_T$ spectra could serve as a handle to disentangle partonic from hadronic sources (breaking parton-hadron duality). For details see~\cite{joao06}.

{\bf Intermediate mass region.} To understand the origin of the excess measured by NA50 in the IMR,  NA60 has measured the offset of the muon tracks with respect to the main interaction vertex. This distribution has been weighted by the inverted error matrix from the vertex fit and the muon extrapolation (the combinatorial background was subtracted by event mixing). The resulting distribution is composed of two components, the Drell-Yan (i.e. prompt) and the open charm (off vertex) events. To evaluate each contribution, their shapes were obtained from Pythia (details are given in \cite{ruben06}); then, the offset distribution was fitted as a superposition of prompt and off-vertex contributions. The fit parameters are the coefficients by which each contribution should be scaled in order to describe the data.

\begin{figure}[htb]
\vspace{-3.mm}
\resizebox{0.50\textwidth}{0.22\textheight}{%
\includegraphics*{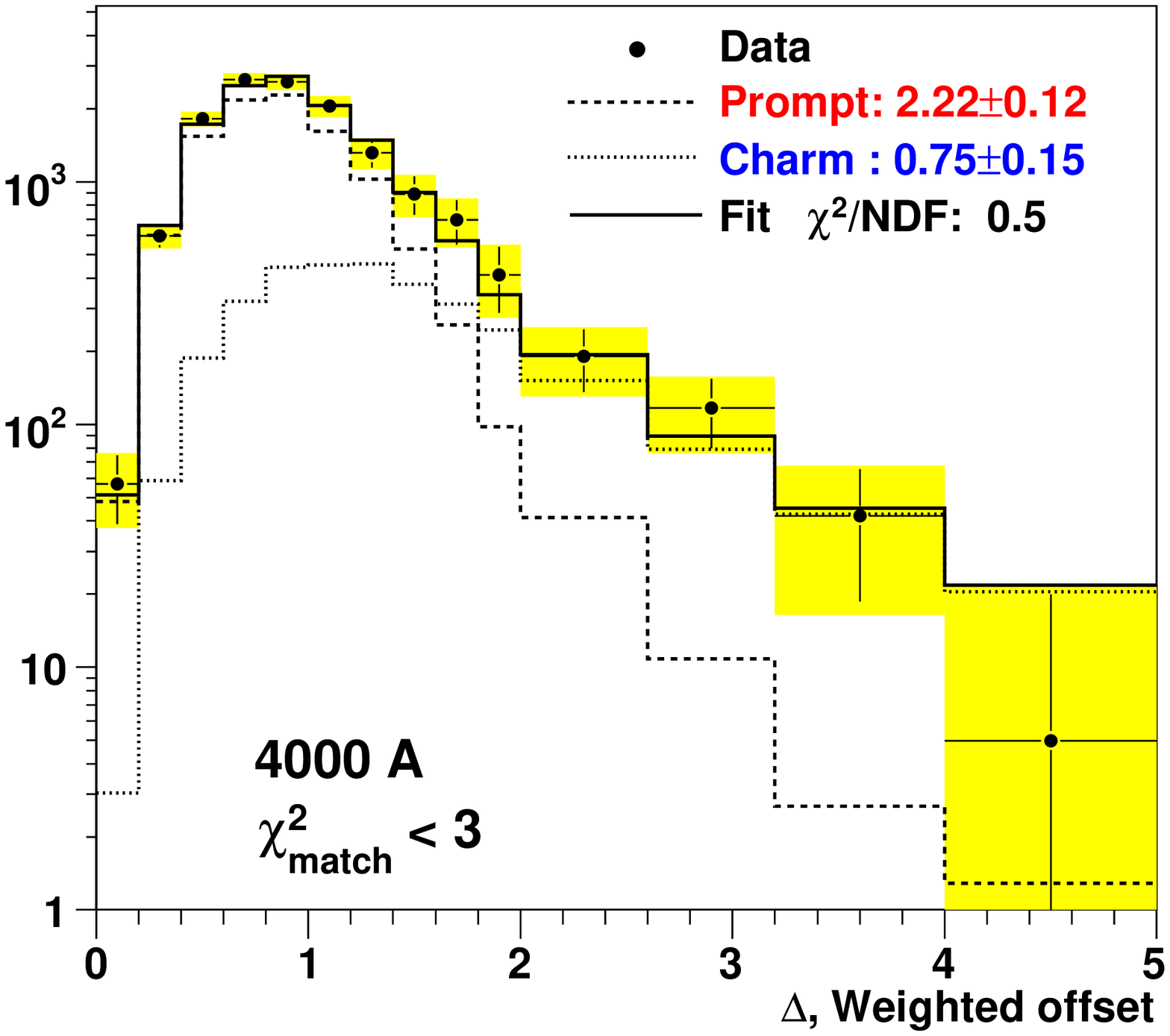}}
\resizebox{0.50\textwidth}{0.22\textheight}{%
\includegraphics*{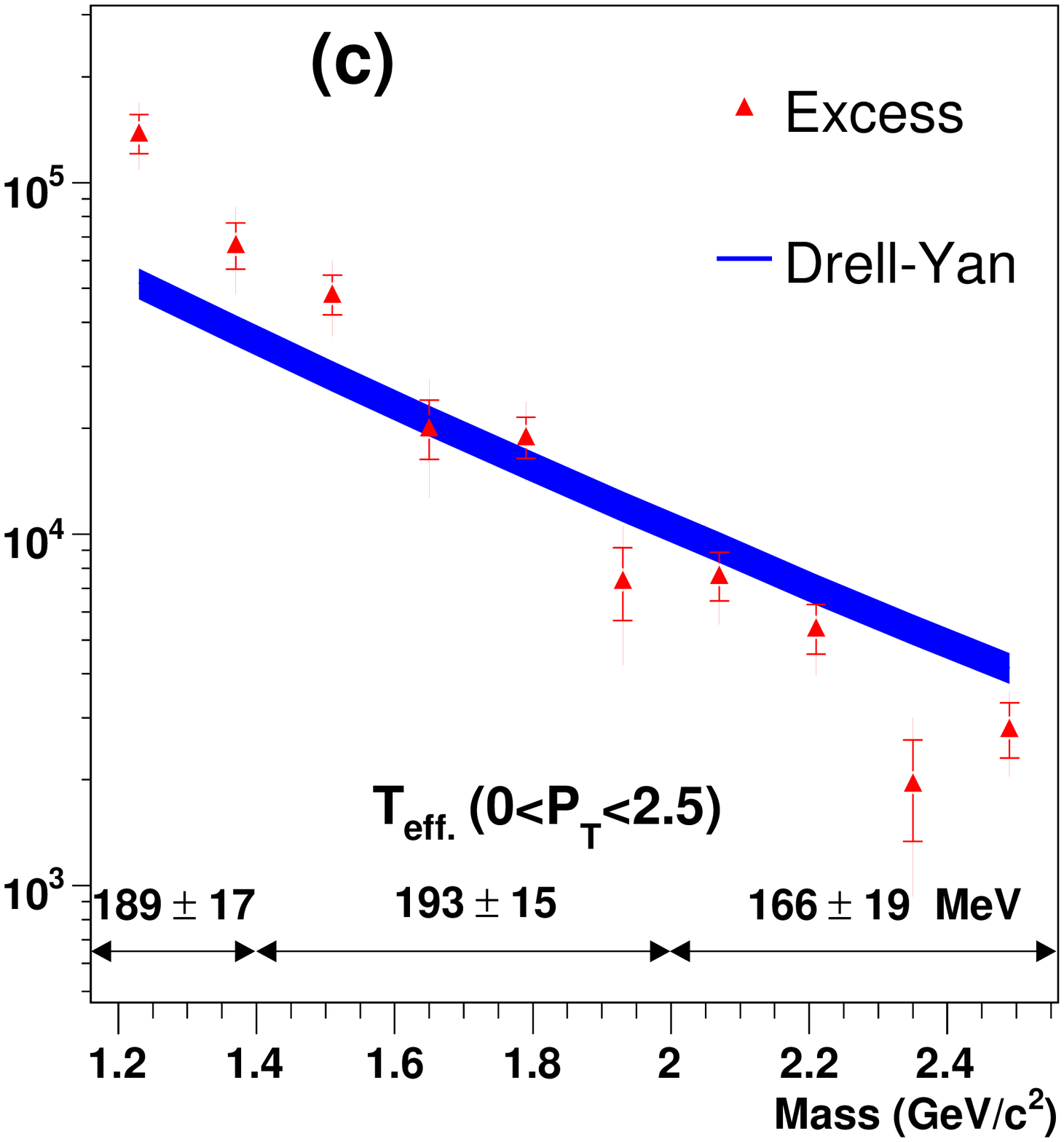}}
\vglue -4mm
\caption{Left panel: fit to the weighted offset distribution where both the prompt and the open charm yields are free parameters  ($1.16<m_{\mu\mu}<2.56$). Right: mass spectra of Drell-Yan and excess. The effective temperature of the excess fitted in the $0<p_T<2.5$ range for the mass bins of
$1.16-1.4$, $1.4-2.0$ and $2.0-2.56$ GeV$/c^2$ is also shown.}
\label{fig2}
\vspace{-3.mm}
\end{figure}

The fit fails to reproduce the data if the prompt yield is forced to 1.1 the expected Drell-Yan yield, and the open charm is left free. In the left panel of fig.\ref{fig2} both contributions are left free: in this way, an accurate representation of data is achieved. The fit parameters indicate that the prompt contribution is two times larger than expected: we can then conclude that the excess is due to a prompt source.

The naive hypothesis that the excess could be due to an increased Drell-Yan yield is ruled out by the $p_T$ dependence of the ratio between the excess and the Pythia-generated Drell-Yan (after acceptance correction). It ranges from 3 at low $p_T$, to 0.5 at high $p_T$, suggesting that the excess is qualitatively different from Drell-Yan. This difference is confirmed by the comparison of the excess and Drell-Yan mass spectra, shown in the right panel of fig.\ref{fig2}. The temperatures shown in the figure are obtained from fits in the range $0<p_T<2.5$~GeV/$c$: the systematic error on the temperature determination has been evaluated from fits performed in different $p_T$ ranges and turns out to be of the order of 10~MeV (see also~\cite{ruben06}).

\begin{figure}[htb]
\resizebox{0.50\textwidth}{0.22\textheight}{%
\includegraphics*{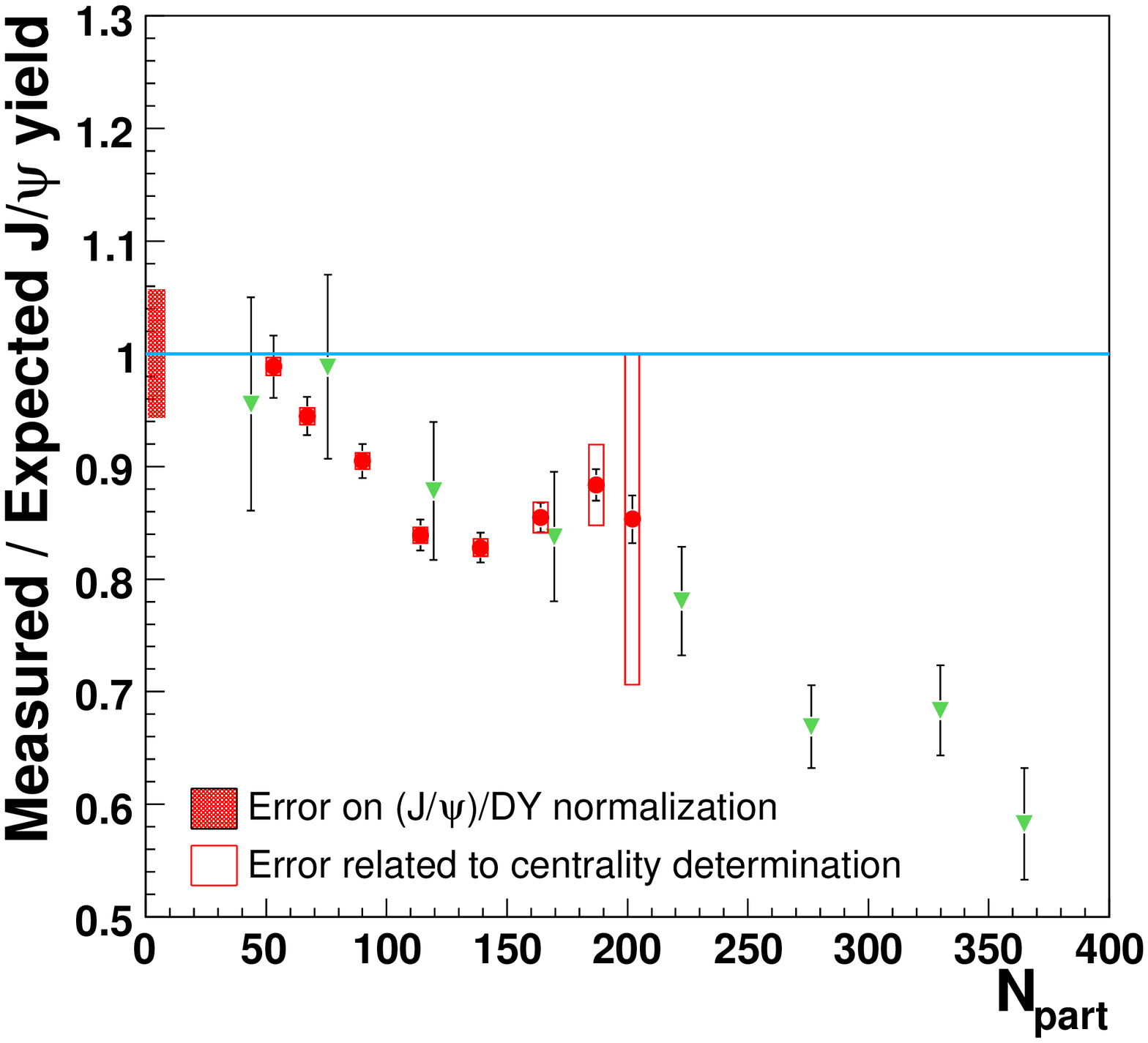}}
\resizebox{0.50\textwidth}{0.22\textheight}{%
\includegraphics*{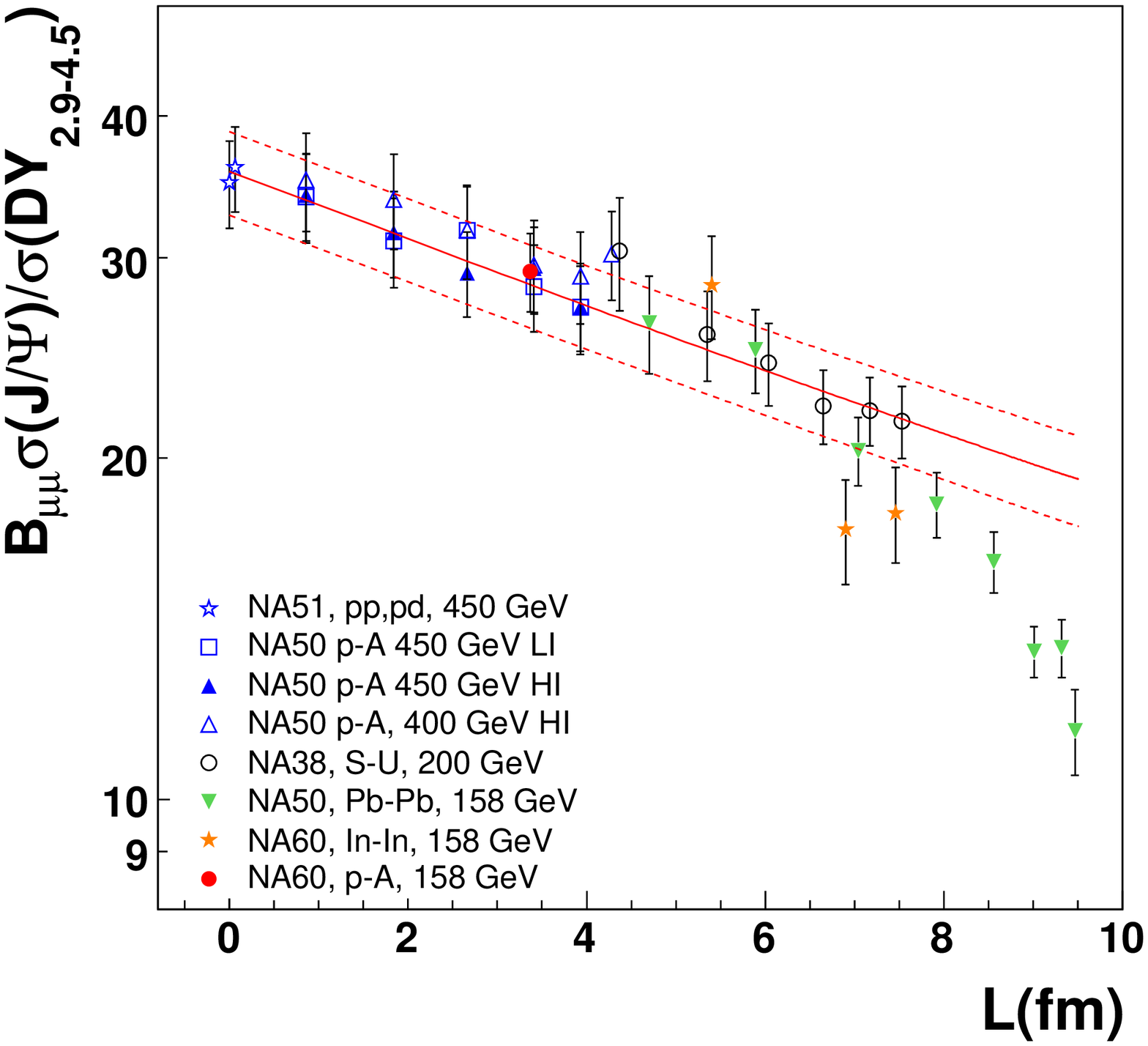}}
\vglue -4mm
\caption{Left: comparison between the In-In (NA60: square points) and
Pb-Pb (NA50: triangle points) suppression patterns. Right: Compilation of the $\sigma_{J/\psi}/\sigma_{DY}$ values
measured in p--A and A-A collisions at the SPS, rescaled,
when necessary, to 158~GeV$/c.$ The lines indicate the results of a
Glauber fit to the p--A data and the size of the error. The full circle
indicates the preliminary NA60 result for p--A collisions at 158~GeV$/c$.}
\label{fig3}
\end{figure}

{\bf J/$\psi$ suppression in In-In.} To measure the amount of anomalous suppression, the J/$\psi$ yield has to be compared to a reference process. Traditionally, the Drell-Yan yield above 4~GeV/$c^2$ has been used, but this choice increases the statistical error due to the DY limited statistics. To overcome this limitation and fully exploit the J/$\psi$ sample, NA60 has calculated the reference spectrum expected in case of normal suppression only: the relative normalization between the calculated and measured spectra is set to the value resulting from the $\sigma_{J/\psi} / \sigma_{DY}$ analysis (see~\cite{enrico06}). 
The resulting suppression pattern is shown in fig.\ref{fig3} (left), compared to the results published by NA50 in Pb--Pb collisions: both sets of data depart from the normal nuclear absorption line for $50 < N_{part} < 100 $, suggesting that $N_{part}$ could be well suited as scaling variable between different colliding systems. The data have been compared to theoretical predictions tuned for NA60 In--In collisions: none of them was able to reproduce data, even if the magnitude of suppression is reasonably reproduced (further details in \cite{enrico06}). 

To evaluate correctly the anomalous suppression, the normal nuclear suppression (which can be extracted from p--A data) must be accurately known. However, up to 2004, p--A data were available only for proton energies of 400 and 450~GeV/$c$: this implied that the p--A results had to be rescaled to the energy of nuclei beams (158~AGeV/$c$), under the assumption that $\sigma_{abs}$ would not change with the beam energy. In order to eliminate uncertainties associated with this assumption, NA60 has taken p--A data at the same energy of the Indium beam on a variety of targets (Be, Al, Cu, In, W, Pb and U). Up to now, a preliminary estimate of $\sigma_{J/\psi} / \sigma_{DY}$ has been obtained by averaging over the different targets. In fig.~\ref{fig3} right, it is shown a comparison of this result to the previous measurements (rescaled when necessary) as a function of the mean length of nuclear matter traversed by the J/$\psi$. The NA60 average point corresponds to a length $\langle L \rangle = 3.4$~fm and falls along the interpolating band, corroborating the correctness of the rescaling procedure and confirming the anomaly of the J/$\psi$ suppression with respect to a pure nuclear absorption scenario.


\section*{References}
\begin{thebibliography}{99}
\bibitem{Usai:2005zh} G. Usai et al. (NA60) Eur. Phys. J. {\bf C43}, 415
  (2005).
%
\bibitem{prl} R. Arnaldi et al. (NA60), Phys. Rev. Lett. {\bf 96}
(2006) 162302.
%
\bibitem{lowmassHP06} S.~Damjanovic {\em et al.} (NA60),
Nucl. Phys. A, 783 (2007) 327.

\bibitem{Brown} G.E. Brown, M. Rho, Phys. Rept. {\bf 363}, 85 (2002) and references therein.
%
%
\bibitem{Rapp1} R. Rapp, J. Wambach,
Adv. Nucl. Phys. {\bf 25}, 1 (2000) and references therein.
%
%
%
%
%
%
%

%
\bibitem{joao06} J.~Seixas {\em et al.} (NA60), to be published in Quark Matter 2006 proceedings, and references therein.
%
\bibitem{NA49systematics} D.~Rohrich,  J.
  Phys. G, 27 (2001) 355.
%
\bibitem{ruben06} R.~Shahoyan {\em et al.} (NA60), to be published in Quark Matter 2006 proceedings, and references therein.
%
\bibitem{enrico06} E.~Scomparin {\em et al.} (NA60), to be published in Quark Matter 2006 proceedings, and references therein.
%
%
%
%
%

%
%
%
%






\end{thebibliography}
\end{document}